# Quantification of Entanglement from Magnetic Susceptibility for a Heisenberg Spin 1/2 system

Tanmoy Chakraborty,[1] Harkirat Singh,[1] Diptaranjan Das,[1] Tamal K. Sen,[1] Swadhin K. Mandal,[1] Chiranjib Mitra.[1*]

[1]Indian Institute of Science Education and Research (IISER) Kolkata, Mohanpur Campus, PO: BCKV CampusMain Office, Mohanpur - 741252, Nadia, West Bengal, India.

*chiranjib@iiserkol.ac.in

**ABSTRACT:** Temperature and magnetic field dependent magnetization and extraction of entanglement from the experimental data is reported for dichloro(thiazole)copper(II), a Heisenberg spin chain system. The magnetic susceptibility vs. temperature plot exhibits signature of infinite spin chain. Isothermal magnetization measurements (as a function of magnetic field) were performed at various temperatures below the antiferromagnetic (AFM) ordering, where the AFM correlations persist significantly. These magnetization curves are fitted to the theoretically generated Bonner-Fisher model. Magnetic susceptibility can be used as a macroscopic witness of entanglement; using which entanglement is extracted from the experimentally measured magnetic data.

## 1. Introduction

The study of entanglement in quantum systems is an area of intense research primarily due to its usefulness as a fundamental resource in quantum computing and quantum communication [1]. Idea of entanglement lies in the fundamentals of quantum mechanics and was first discussed by Einstein in 1935 [2]. Entanglement refers to quantum correlation between quantum mechanical particles that is entirely nonclassical in nature and hard to visualise within the purview of classical world. Entanglement is inherently a non-local phenomena where the two entangled particles violate Bell's inequality [2]. In the past couple of decades it has been studied extensively both theoretically and experimentally in the context of quantum information science. For instance, it has important applications in the area of a quantum information processing [3] and in the research of quantum cryptography [4]. Hence study of entanglement in condensed matter systems is relevant while dealing with many body systems which have quantum mechanical correlation between its constituent particles [5].

Measures of entanglement in many body condensed matter system is an area of active research. It has been reported that entanglement is quantifiable and it is possible to measure how much entanglement exist in some solid state system by using Entanglement Witnesses (EW) [6, 7, 8 and 9]. EW is an observable that takes up a non-positive value for separable states and positive value otherwise, indicating entanglement. Extraction of entanglement from experimental data was first done experimentally by carrying out bulk measurement on an insulating system $LiHo_xY_{(1-x)}F$ [10]. Consequently, various measures of entanglement have been proposed. In their work, Brukner et al [6] discuss how entanglement in certain systems can have signatures even in the thermodynamic limit. It is already reported that macroscopic thermodynamic observables like susceptibility, specific heat can capture nonlocal correlations amongst the microscopic constituents (spins) in solid state systems. This could be used to extract and quantify entanglement in these systems using the fact that non-separability leads to violation of uncertainty bounds obtained by assuming separability [6, 11]. In this work, magnetic susceptibility is used as an entanglement witness and quantification of entanglement is done by extracting information from the macroscopic magnetic susceptibility data.

One dimensional exchange coupled magnetic systems have been studied extensively both theoretically and experimentally. However studying different spin chain systems from the perspective of quantum information science has been a field of great interest in the last few decades. One of the prospective applications of spin ½ chains where the spins are entangled, is that they can be used as a medium for quantum communication [12, 13]. Spin ½ chain systems provides an excellent platform for exploring about entanglement content. This is because due to the built in symmetry in the Hamiltonian of these systems, exact numerical diagonalization is possible [14]. The variation of entanglement with magnetic field and temperature has been shown experimentally for a spin ½ systems with Heisenberg interaction [15]. In absence of any external field the ground state of an antiferromagnetic system is entangled, because the order parameter (which is staggered magnetization for systems with J>0) doesn't commute with the Hamiltonian thus exhibiting quantum fluctuations which prevent long range ordering.

We report detailed magnetic study and extraction of entanglement from experimentally observed magnetic susceptibility data for dichloro(thiazole)copper(II), a Cu based polymer system [16, 17]. It has been confirmed from energy calculations that antiferromagnetic state is a stable ground state for this system [16]. Magnetic centre for this system, resides in the Cu atom and each molecule bears one unpaired spin. The structure consists of square-planer units arranged in arrays. These units are equatorially linked along a particular axis by chloride bridges to form infinite linear chains [16, 17]. Earlier study has shown, [17] that for this system, the unpaired spins interact with other neighbouring spins along a particular direction by isotropic Heisenberg interaction. This periodic arrangement of spins along a particular direction enables us to represent this system like a spin-1/2 chain which is translationally invariant. Estes *et.al*. [17] ensured 1D magnetic interaction of this compound by fitting its susceptibility vs. temperature data to Bonner-Fisher model in low temperature regime although it has small interchain interaction at relatively higher temperature. To study the entanglement in this system, one needs to be concerned only in the low temperature regime; therefore one can consider this as a 1D system. In this system, the intrachain interaction between the nearest neighbour is antiferromagnetic with the same coupling strength. The general form of the spin chain Hamiltonian for a finite number of spins with nearest neighbour interaction is

$$H = 2J \sum_i \left[ \alpha S_i^z S_{i+1}^z + \beta (S_i^x S_{i+1}^x + S_i^y S_{i+1}^y) \right] \qquad (1)$$

J is the exchange integral and $S_x$, $S_y$, $S_z$ are the components of the total spin S along x, y and z direction respectively. For α=1 and β=0 Ising model is obtained, for α=0 and β=1 the spins are restricted in the X-Y plane and Heisenberg X-Y model is realized. When α=β=1 the Hamiltonian takes the form of isotropic Heisenberg model. For the system to be antiferromagnet, J>0. Bonner and Fisher [18] carried out calculation on isotropically interacting spin chain systems containing spins from 3 to 11 and extrapolated it for infinite chain with a good agreement. Hall [19] found an excellent analogy of the numerical Bonner-Fisher model for zero field susceptibility to the expression given below (equation 2) and fitted efficiently to the theoretically calculated Bonner-Fisher data,

$$\chi \approx \frac{Ng^2 \mu_B^2}{k_B T} \cdot \frac{0.25 + 0.14995X + 0.30094X^2}{1.0 + 1.9862X + 0.68854X^2 + 6.0626X^3} \qquad (2)$$

where, $X = J/K_B T$ and the other symbols have their usual meaning. Magnetic susceptibility is experimentally measured as a function of temperature and fitted to the above mentioned analytical expression. We obtained a good fit consistent with the previous literature [17]. Subsequently, magnetization isotherms with varying magnetic field are taken at different temperatures. These experimentally measured magnetic datasets are used to quantify the amount of entanglement in this spin chain system.

## 2. Materials and Methods

Thiazole ($C_3H_3NS$), Copper chloride dyhydrate ($CuCl_2$, 2H2O) and absolute Ethanol, purchased from SIGMA ALDRICH with purest grade, were used as starting reagents. We followed the synthesis route as mentioned in the ref [17] and obtained nice blue crystals. The magnetic measurements of $Cu(tz)_2Cl_2$ were performed in a Quantum Design MPMS (Magnetic Property Measurement System). The static magnetic susceptibility vs. temperature data were collected in a temperature range of 2K to 24K. Subsequently, magnetization isotherms as a function of magnetic field were taken at various temperatures from 2K to 20K. The field was varied from 0 to 7 Tesla.

## 3. Results and Discussion

We have used magnetic susceptibility as a macroscopic entanglement witness to analyze the results of magnetic measurements for a spin 1/2 antiferromagnetic chain. Experimentally measured magnetic susceptibility data as a function of temperature is shown in Fig. 1(a) in the temperature range of 2 K to 24 K. The data is fitted to Bonner-Fisher model using the analytical expression given in eqn. (2) taking into account the boundary effects, which results paramagnetic contribution to the susceptibility data. Exchange coupling constant J was used as fitting parameter and the value of J that was obtained after fitting is 5.6 K, which is in good agreement with the value reported in the prervious literature [17]. An entanglement witness (EW) is able to certify if a certain system is entangled or not and it was first introduced by Horodecki *et.al.* [20]. An observable $W$ for a state $\rho$ is an EW, if $Tr(\rho W) \leq 0$ when $\rho$ is separable and $Tr(\rho W) \geq 0$ otherwise [21]. Magnetic susceptibility is a macroscopic entanglement witness, which for spin ½ chain systems consisting N spins ½ particles is given as

$$EW(N) = [1 - \frac{6k_B T \chi}{g^2 \mu_B^2 N}] \qquad (3)$$

Where $\chi$ is experimental susceptibility and the other symbols have their usual meanings. EW provides necessary and sufficient condition for existence of entanglement. For a state to be entangled, the expectation value of EW at that particular state should be within a certain bound; although for expectation value exceeding the bound does not assure separability [5]. In our case, the condition $1 > \frac{6k_B T \chi}{g^2 \mu_B^2 N}$ leads to the existence of entanglement in the system. The entanglement boundary is shown by the dotted curve (Fig. 1.a). The entangled region is represented by the region towards the left of the dotted curve and this demarcation is governed by the condition "$1 > \frac{6k_B T \chi}{g^2 \mu_B^2 N}$". This entanglement boundary intersects the susceptibility curve at 12 K. That means the system exhibits entanglement up to 12 K. This is the critical value of the temperature above which the system may or may not remain entangled. The entanglement exists up to a temperature (12 K in this case) where antiferromagnetic correlations persist and this value of temperature is more than the magnetic ordering temperature. The magnetic ordering temperature corresponds to the peak in the susceptibility curve, which is 7.5 K in this case. The magnetic ordering in this case is not strictly a long range ordering akin to a classical magnetic system. The presence of spin fluctuations mentioned above, presents any kind of long range order. The value of EW is extracted from experimentally measured temperature dependent susceptibility data using the mathematical expression for EW (eqn. 3). The extracted value of entanglement is plotted as a function of temperature and is shown in Fig.1 (b). Since we are only concerned within estimation of EW in the region below the critical temperature, all EW values above this particular temperature are considered to be zero. One can see that EW vanishes at 12 K, which is the critical temperature mentioned above.

Isothermal magnetization measurements were performed for Cu(tz)$_2$Cl$_2$ as a function of magnetic field for various temperatures. These measurements are concentrated in the temperature regime where magnetic ordering occurs, is 7.5 K in this case. Magnetization isotherms are taken in the temperature range 3 K to 24 K. A 3D plot (Fig. 2) is generated from experimental data which explicitly shows the variation of magnetization with magnetic field and temperature for this system.

The magnetization isotherms at 3 K and 5 K are compared with the theoretical Bonner-Fisher model of spin chain for 10 spins. Bonner-Fisher numerically generated isothermal magnetization for N=10 number of isotropically interacting spins and assured the convergence for N tending to infinity. We have numerically calculated theoretical magnetization for 10 spins as a function of field and plotted in the same graph with the experimental ones at corresponding temperatures. In this programme we have used J=5.6K (obtained from the susceptibility vs. temperature fitting). In Fig. 3 one can clearly see, that the theoretical curves are in good agreement with the experimental ones.

Till now the quantification of entanglement has been done from susceptibility vs. temperature data. In this part of the paper we will use the experimental magnetization isotherms at different temperatures to extract the amount of entanglement. Magnetic susceptibility ($\chi$) is defined as the field derivative of magnetization M ($\chi = Lt_{B \to 0} \frac{\partial M}{\partial B}$), where B is the magnetic field. Hence the magnetization isotherms as function of field enable us to capture the variation of susceptibility and hence entanglement with temperature and magnetic field. We have derived temperature dependent susceptibility at different fields using the above formula. We have used these susceptibility datasets as EW and plotted the quantified experimental entanglement as a function of magnetic field and temperature. This is shown in the 3D plot given in Fig. 4. It can be seen that Entanglement is decreasing with increasing temperature and comes to zero at 12 K which is compatible with the analysis done on the experimental susceptibility vs. temperature data. From this plot one can see that the entanglement at low temperature does not change significantly with the change in magnetic field. This is owing to the strong coupling strength between the spins in the spin chain. Upon application of strong magnetic field, in a mixed state the contribution to the statistical weight of ferromagnetic states increases, thereby reducing the entanglement in the system. This is also tantamount to formation of triplet states at the expense of singlet states. However, there are two competing energy scales, the exchange coupling 'J' and the external magnetic field 'B'. To cause significant pair breaking, one needs to apply a significantly higher magnetic field than was possible in our case. Thus we couldn't affect any substantial change in entanglement upon application of magnetic field.

4. **Conclusion**

With the advent of quantum information processing there is a renewed interest in spin chains especially where the ground state is antiferromagnetic. This is because such a ground state is entangled and for a spin ½ system, this entanglement can be well characterized both theoretically and experimentally. We considered one such spin ½ chain ($Cu(tz)_2Cl_2$) and characterized the entanglement content from magnetic susceptibility data using an entanglement witness. We also studied its variation both as a function of temperature and field. The application of field as well as increase in temperature causes a mixing in the states and thus the ground state is no longer an entangled pure state. Thus it is seen that the entanglement reduces upon increasing the temperature. However, the magnetic field applied isn't large enough to cause significant amount of spin flip or pair breaking to reduce the entanglement at the lowest temperature and this is reflected in our analysis. The entanglement obtained using the witness operator is the average entanglement between any two spins [11].

In addition to this we have also fitted our magnetic susceptibility and magnetization data to the "Bonner-Fisher" model [18] to ascertain that this compound is an antiferromagnetic spin ½ chain system. Our fits were excellent indicating that this is indeed an archetypal antiferromagnetic spin ½ system and a very suitable candidate for the study of entanglement. This characterization of entanglement in spin chains can be useful in identifying suitable systems for quantum communication and also enable them to be useful as quantum networks connecting two quantum gates [12, 13].

**Acknowledgments**

The authors would like to thank the Ministry of Human Resource Development (MHRD), Government of India, for funding.

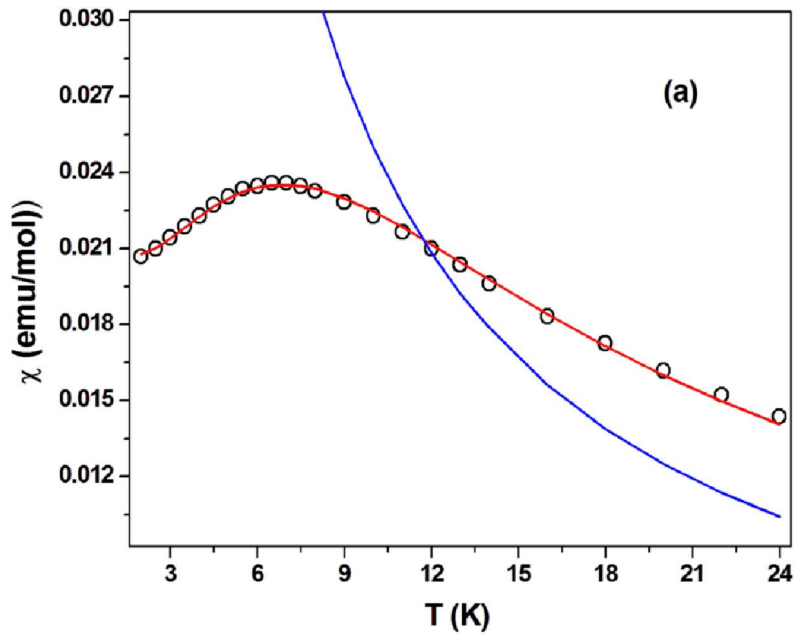

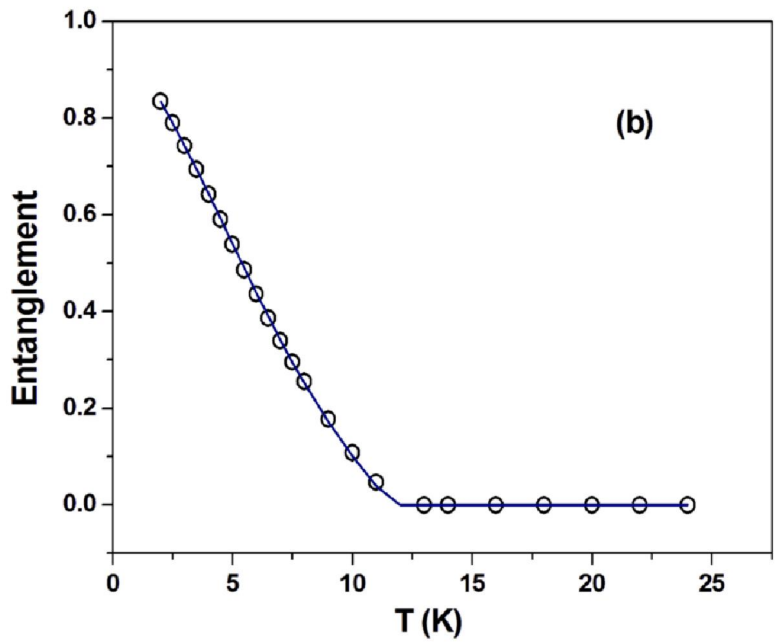

**Fig. 1.** (a) Susceptibility data (circles) of Cu(tz)$_2$Cl$_2$ and fit (solid line) to Eq. 2, the isotropic Heisenberg chain model. The dotted line demarks the entangled regime (see text). (b) Extraction of entanglement witness from the susceptibility data of Cu(tz)$_2$Cl$_2$ using Eq. 3.

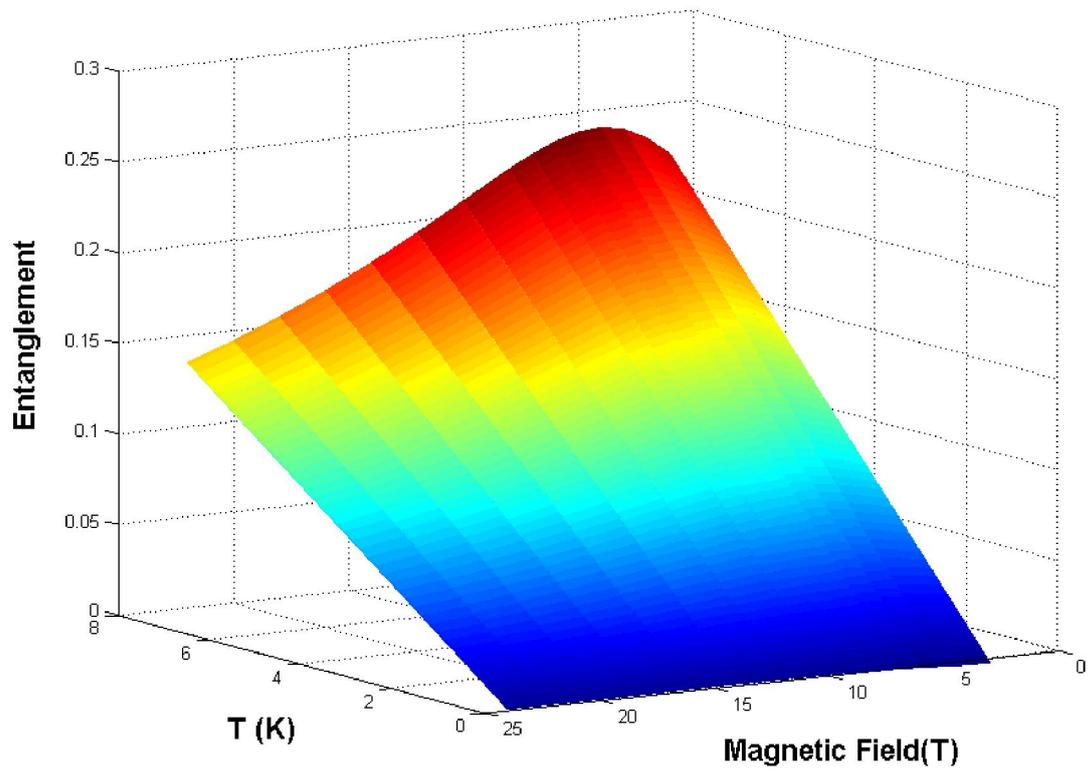

**Fig. 2.** 3D plot showing Magnetization, magnetic field and temperature along the three axes for Cu(tz)$_2$Cl$_2$.

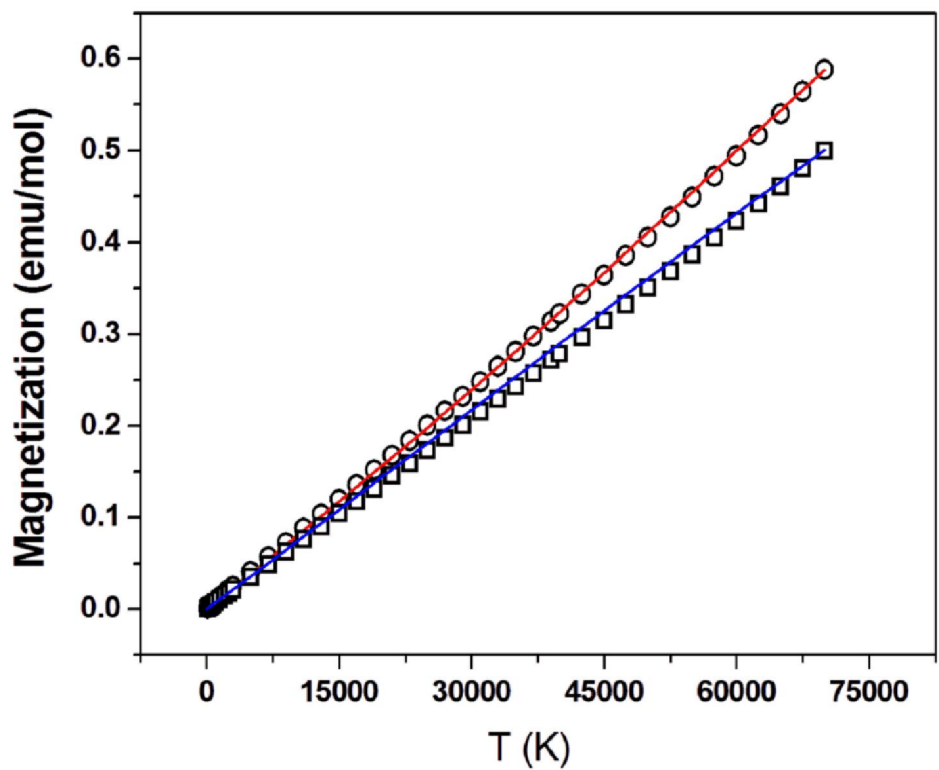

**Fig. 3.** Experimental data of magnetization collected at T = 3 K and 5 K for Cu(tz)$_2$Cl$_2$ (open circles and open squares respectively) as a function of magnetic field and the theoretical curve at T = 3 K and 5 K (solid red line and solid blue line) derived using the Bonner-Fisher model.

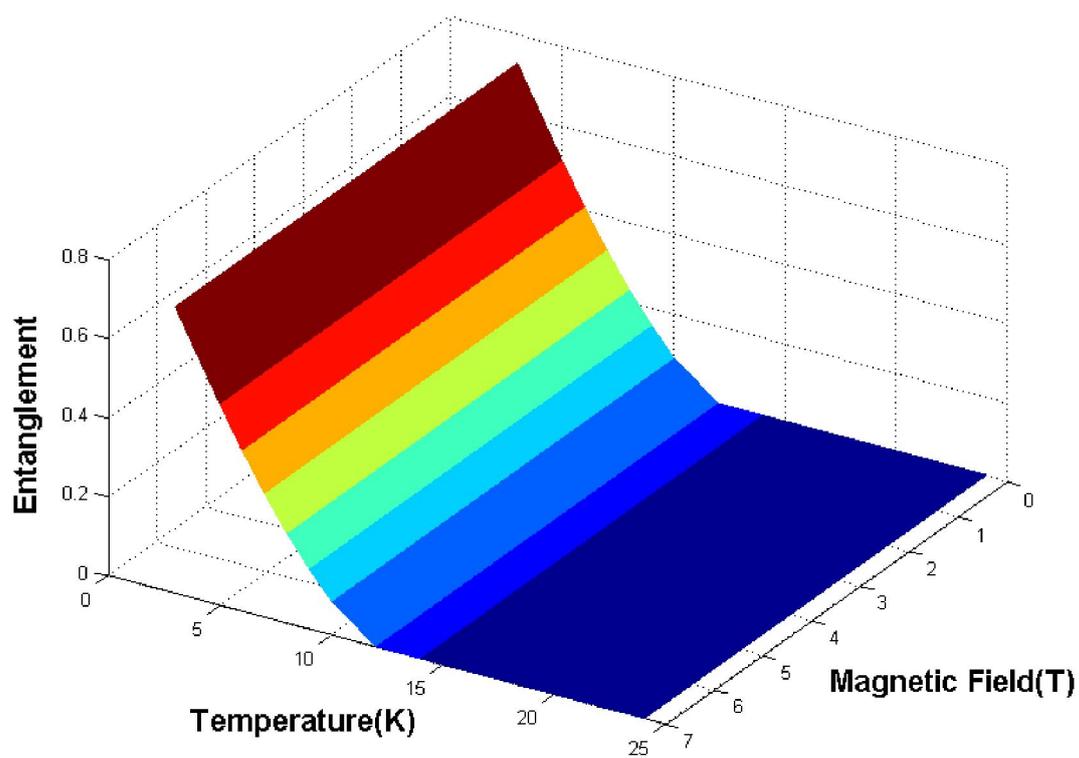

**Fig. 4.** Experimental value of concurrence as a function of magnetic field and temperature for Cu(tz)$_2$Cl$_2$. The magnetic field values are in Tesla and the temperature is in Kelvin.